\documentclass[12pt]{article}

\usepackage{amsmath,amsfonts,amssymb}

\def\1barra{1\! \hskip -1.1pt {\rm l}}

\def\0barra{{\rm O} \!\hskip -3.7pt {\rm l} }

\title{ Berry's Phase in the Presence of a
Dissipative Medium }

\author{ K.M. Fonseca Romero$^{(1)}$,  A.C. Aguiar Pinto$^{(2)}$,
          \\
             and    M. T. Thomaz$^{(2)}$
 \\
\\
\baselineskip =10pt
{\small\it $^{(1)}$Universidad Nacional, Facultad de Ciencias,}
\vspace{-0.2cm} \\
{ \small \it   Departamento de  F\'\i sica, Ciudad Universitaria,}
\vspace{-0.2cm} \\
{ \small \it Bogot\'a, Colombia} \\
{ \small \it $^{(2)}$ Instituto de F\'\i sica - Univ. Federal
                      Fluminense
\vspace{-0.2cm}}\\
{\small \it Av. Gal. Milton Tavares de Souza s/n.$\!\!^\circ$,
             \vspace{-0.2cm} }\\
{ \small \it CEP: 24210-340, Niter\'oi, R.J.,   Brazil }  \\}

\date{  }

\begin{document}

\maketitle

\begin{abstract}

We consider the  spin 1/2 model coupled  to a slowly varying
magnetic field in the presence of a weak damping represented  by a
Lindblad-form operators. We show  that Berry's geometrical phase
remains unaltered by the two dissipation mechanism
considered.  Dissipation effects are
twofold: a shrinking  in the modulus of the Bloch's vector, which
characterizes  coherence loss and a time dependent (dissipation
related) precession angle. We show that the line broadening of the
Fourier transformation of the components of magnetization is only
due to the presence of dissipation.

\end{abstract}

\vfill

\baselineskip=12pt



\newpage

\baselineskip=18pt

\section{Introduction}

The existence of geometric phases in quantum systems, ever since its
discovery by Berry \cite{berry}, has attracted considerable interest both
from  the theoretical and experimental viewpoints. Several applications
 of this phenomenon in different areas of physics have also been
studied\cite{experimental,experimental1,experimental2}.
Comparatively much less work has been devoted to the question of the
dynamical evolution of such systems in the presence of a weakly
dissipative medium. Basically the available results can be summarized as
follows: nonhermitian operators lead to a modification
of Berry's phase\cite{physlet_a,physlet_a1, physlet_a2}, stochastically evolving magnetic
fields produce both energy shift and broadening\cite{gaitan},
phenomenological weakly dissipative Liouvillians alter Berry's phase
by introducing an imaginary correction\cite{pra} or causing damping and mixing of
the density matrix elements\cite{physrev_a}.
Ellinas {\it et al.} \cite{pra} obtain their results studying  the eigenmatrices of the
complete Liouvillian  superoperator in a time-independent basis, while
Gamliel and Freed\cite{physrev_a}, find
 closed formal expressions for the density matrix representation
in the instantaneous frame of the hamiltonian in the adiabatic limit,
and in the  weak dissipation approximation. However, even in this regime the results
 in the cases studied can only be extracted through approximations
(in addition to the adiabatic approximation) or numerical
computation.

In the present work, we consider the celebrated example of a spin
1/2 particle coupled to a slowly evolving magnetic field in the
presence of a weakly dissipative medium as represented by a
Lindblad form superoperator\cite{lindblad,lindblad1,lindblad2},
incorporated in a convenient and physically motivated frame. In
the absence of dissipation, a simple geometrical interpretation of
the results
 emerges in terms of both the Bloch's vector\cite{pra} and the phase vector\cite{muller}.
The geometrical phase appears as a ``delay'' or an ``advance''
 in the precession period of the Bloch's vector
 with respect to the period dictated by the
magnetic field's frequency. The precession of this vector occurs
at a fixed angle with respect to a fixed axis about which the external
magnetic field precesses, as usual. We will introduce the dissipation
via a semi-group type dynamics, such that dissipation
does not alter the precession frequency. However, the $x$, $y$ and $z$
components of Bloch's vector are altered in different ways by
nonunitary effects. In particular,  the modulus of Bloch's
vector will shrink. The relation between the reduction of 
Bloch's vector modulus and the loss of coherence has been
explored by Stodolsky and collaborators\cite{harris, stodolsky,raffelt}.
In order to make it explicit, we introduce the linear entropy
(or idempotency defect) as a measure of purity loss\cite{prl}

\begin{equation}
\delta (t) = 1 - Tr \; \rho^2 (t)
      \label{1}
\end{equation}

\noindent and write

\begin{equation}
\rho (t) = \frac{1}{2} \left( \begin{array}{cc}
1+S_z(t) & S_x(t) -i S_y(t) \\ S_x(t) +i S_y(t) & 1 - S_z(t)
     \end{array}\right) =
 \frac{1}{2} (  \1barra
 +\vec{S}(t) \cdot \vec{\sigma}),
                 \label{2}
\end{equation}

\noindent  where $\vec {S}(t)$ is the Bloch's vector. We get

\begin{equation}
\delta (t) = \frac{1 - | \vec{S} (t) |^2}{2},
        \label{3}
\end{equation}

\noindent which confirms that the shrinking of the
modulus of $\vec{S} (t)$ is a measure of
coherence loss.

In order to discuss dissipative effects on interference patterns due to
Berry's geometrical phase the
natural framework is to consider the evolution of the density matrix,
which completely  characterizes the interference effect due to
the  different geometric phases acquired by the eigenvectors of the hamiltonian
 of the system, even in the absence of dissipation\cite{pra}. Of course, this time we add
 a nonunitary Liouville operator contribution to the dynamics.

In section 2 we present the two level quantum system of a spin
1/2 in the presence of an external magnetic field precessing with
constant angular velocity around a fixed axis. The master
equation of the model is written in what we call the diagonal
frame where Lindblad superoperators are introduced to describe
the nonunitary part of the Liouville operator. In subsection 2.1
we consider the adiabatic limit of the quantum system in thermal
equilibrium with a reservoir of electromagnetic fields and in
subsection 2.2 we consider the case of a dephasing process. In
both cases, we treat the coupling of the quantum system with its
environment in the weak regime. In section 3 we show how the
geometric phases and the dissipation effects acts on the $z$
component of the magnetization of the spin 1/2 coupled to a
reservoir of electromagnetic fields at thermal equilibrium. In
section 4 we summarize our conclusions; finally, in appendix A,
we present the non-unitary part of the liouvillians of the models
under consideration in the instantaneous basis of the hamiltonian.

\section{The adiabatic limit of the spin 1/2 model in the \hfill  \break
weak coupling regime}

We consider a spin 1/2 variable (two level model)  coupled to a time dependent
 magnetic field  precessing around the z-axis. The unitary contribution for this
evolution is given by the hamiltonian

\begin{eqnarray}
H_s(t) & = & \mu \vec{\sigma} \cdot \vec{B}(t)   \nonumber \\
%
%
& = &  \mu B \left( \begin{array}{lr}
\cos(\theta) & \sin(\theta) e^{-i\omega t}\\
\sin(\theta) e^{i\omega t} & -\cos(\theta)
\end{array}\right),
     \label{4}
\end{eqnarray}

\noindent written in the basis of the eigenstates of the $z$-component
of the spin, where $B$ is the norm of the external magnetic
field, $\theta$  its azimuthal angle, $\omega$ the
precession frequency and the constant $\mu=\frac{g \mu_B}{2}$, being
$g$   Land\'e's factor and $\mu_B$ the Bohr magneton.
We are using natural units ($c=\hbar=1$). For the sake
of later calculations, it is convenient to define two unitary
transformations: the first one, $R(\omega, t)$, takes us to the rotating
 frame where the hamiltonian is no  longer time dependent; 
  the second one, $D(B, \theta,\omega)$, diagonalizes
the effective hamiltonian (time independent) that drives the dynamics of the
final matrix representation of the density operator. After the first
transformation,
$R(\omega, t) = e^{-\frac{i\omega t}{2} \sigma_z}$,
the density matrix and the hamiltonian read

\vspace{-0.5cm}

\begin{equation}
\rho_R(t) = e^{i\frac{\omega t}{2}\sigma_z} \rho(t)
 e^{- i\frac{\omega t}{2} \sigma_z},
          \label{5.0}
\end{equation}

\noindent and

\begin{equation}
H_R =  \mu B (\sin(\theta) \sigma_x + \cos(\theta) \sigma_z) .
    \label{5}
\end{equation}

\noindent In analogous manner, after the second transformation we get, in
the diagonal frame, the density matrix

\begin{equation}
     \rho_D(t) = {\bf D}^T \, \rho_R (t) \; {\bf D},
             \label{6.0}
\end{equation}

\noindent and the effective hamiltonian

\begin{eqnarray}
   H_D & = &  {\bf D}^T ( H_R - \frac{\omega}{2} \sigma_z ) {\bf D}
                             \nonumber\\
%
%
   &  =  & \left( \begin{array}{lr}
                    \lambda_1 & 0\\
                     0 & - \lambda_1
                \end{array}\right)  ,
   \label{6}
\end{eqnarray}

\noindent where
$\lambda_1 = \sqrt{ \mu^2 B^2 \sin^2(\theta) +
 (\mu B \cos(\theta) - \frac{\omega}{2})^2 }$. The rotation matrix
$ {\bf D} = {\bf D}^T$  is equal to

\begin{equation}
{\bf D} =
\sqrt{\frac{1}{2} - \frac{1}{2\lambda_1} (\mu B \cos(\theta) -
\frac{\omega}{2} ) } \;\sigma_x
+
\sqrt{\frac{1}{2} + \frac{1}{2\lambda_1} (\mu B \cos(\theta) -
\frac{\omega}{2}) }\; \sigma_z.
       \label{6.1}
\end{equation}

\vspace{0.5cm}

 One possible way to add dissipative contributions to
 the above dynamics is to include a Lindblad type superoperator
in the evolution equation in the diagonal frame

\begin{equation}
\frac{d}{dt}\rho_D\left( t\right) = -{i} \left[
\lambda_1 {\sigma }_{z},\rho_D\left( t\right) \right]
+k {\cal L}_D \rho_D (t) , \label{6.2}
\end{equation}

\noindent where $k$ is the dissipation constant. The weak coupling regime
 is characterized by the condition $ \frac{k}{\lambda_1} \ll 1$.

Our aim is isolating, in the density matrix, the effects of
dissipation on interference due to geometric phases. The
representation of the density matrix in the diagonal frame is not
very enlightening for this purpose. This is better realized in
some basis of the instantaneous  eigenvectors of hamiltonian
(\ref{4}), as was done by Gamliel and Freed \cite{physrev_a}.  We
define $\rho_I(t)$, the matrix density in a basis of the
instantaneous eigenvectors of hamiltonian (\ref{4})\cite{nota}.
The relation between  $\rho_I(t)$ and $\rho_D(t)$ is

\begin{equation}
\rho_I(t) = {\bf V}^\dag(t)\,{\bf D}\,  \rho_D(t) \, {\bf D} \, {\bf V}(t),
                 \label{7}
\end{equation}

\noindent where the matrix ${\bf V}(t)$ is equal to

\begin{equation}
 {\bf V}(t) = \left( \begin{array}{lr}
                    \cos(\frac{\theta}{2}) e^{-\frac{i\omega t}{2}} &
                                         - \sin(\frac{\theta}{2}) e^{-\frac{i\omega t}{2}} \\
                     \sin(\frac{\theta}{2}) e^{\frac{i\omega t}{2}} &
                     \cos(\frac{\theta}{2}) e^{\frac{i\omega t}{2}}
                \end{array}\right)  .  \label{8}
\end{equation}

The time evolution of $\rho_I(t)$ is given by

\begin{subequations}\label{9}
\begin{equation}
\frac{d}{dt}\rho_I\left( t\right) = -{i} \left[
\left( \mu B + \frac{\omega}{2} \right) {\sigma }_{z} -  \frac{\omega}{2} \sigma_n(t) ,
                \rho_I\left( t\right) \right]
                           +k {\cal L}_I \rho_I (t)\label{9.1}
\end{equation}

\noindent where
\vspace{-0.6cm}

\begin{equation}
 \sigma_n(t) = \left( \begin{array}{lr}
                    \cos(\theta) &  - \sin(\theta) e^{-i\omega t} \\
                     - \sin(\theta) e^{i\omega t}   & - \cos(\theta)
                \end{array}\right) \label{10}
\end{equation}

\end{subequations}

\noindent and ${\cal L}_I \rho_I (t)$ is obtained from ${\cal L}_D \rho_D (t)$ through
a similarity transformation equivalent to (\ref{7}).

Before we specialize our discussion to any particular liouvillian, we study the
adiabatic limit of eqs.(\ref{9}) for the coupling constant $k$ in
the weak regime. In this
regime, the matrix ${\cal L}_I \rho_I (t)$  is written as a linear superposition of the
elements $\rho_{ij}^{I} (t)$. We remind that the density matrix of a two level model
 must satisfy two conditions: {\it i}) $Tr(\rho_I (t)) =1$  $\;$ and 
 {\it ii}) $\rho_{21}^I(t)= (\rho^I_{12}(t) )^* $. As a consequence of those
 conditions, the density matrix has only two independent elements.
 We take the elements
$\rho_{11}^ {I} (t)$ and $\rho_{12}^ {I} (t)$ as our two independent entries.
The general form for the time equations of those two elements in the weak coupling regime
is

\begin{subequations}  \label{11}
\begin{equation}
\frac{d}{dt}\rho_{11}^I (t) = a_{11}(\omega, k; t) \rho_{11}^I (t) +
      a_{12}(\omega, k; t) \rho_{12}^I (t) + a_{13}(\omega, k; t) \rho_{21}^I (t)
          + b_1(\omega, k; t)    \label{11.1}
\end{equation}

\noindent and

\vspace{-0.6cm}

\begin{eqnarray}
\hspace{-0.3cm}
\frac{d}{dt}\rho_{12}^I (t) & = &a_{21}(\omega, k; t) \rho_{11}^I (t) +
    (-2i\mu B +  a_{22}(\omega, k; t)) \rho_{12}^I (t) + a_{23}(\omega, k; t) \rho_{21}^I (t) +
                                  \nonumber\\
      & & \hspace{2cm} + b_2(\omega, k; t)  .  \label{11.2}
\end{eqnarray}

\end{subequations}

We make the change of variables:

\begin{subequations}
\begin{equation}
\rho_{ij}^I (t) \equiv e^{-i(E_i-E_j)t} \;\; \tilde{\rho}_{ij}(t)
\label{12.1}
\end{equation}

\noindent with $E_1 = \mu B$ and $E_2 = -\mu B$. We may disclose  the
time scale $T$ in the differential equations by introducing 
the following transformation upon the time
parameter\cite{born,japones}:

\begin{equation}
 s\equiv \frac{t}{T},
\end{equation}

\end{subequations}

\noindent where $T=\frac{2\pi}{\omega}$.

With the new variables, eqs.(\ref{11}) become

\vspace{-0.5cm}

\begin{subequations}  \label{13.0}
\begin{eqnarray}
\frac{d}{ds}\tilde{\rho}_{11} (s)  &= & T a_{11}(\omega, k; s) \tilde{\rho}_{11} (s)
   + T  a_{12}(\omega, k; s)  e^{-2i\mu BTs} \tilde{\rho}_{12}(s)  +\nonumber \\
         &  + & Ta_{13}(\omega, k; s) e^{2i\mu BTs} \tilde{\rho}_{21} (s)
          + T b_1(\omega, k; s),    \label{13.1}    \\
\frac{d}{ds}\tilde{\rho}_{12} (s) & = &
 T a_{21}(\omega, k; s) e^{2i\mu BTs} \tilde{\rho}_{11} (s)
    + T   a_{22}(\omega, k; s) \tilde{\rho}_{12} (s)  +  \nonumber\\
        & + & T a_{23}(\omega, k; s)  e^{4i\mu BTs} \tilde{\rho}_{21} (s)
            + T  b_2(\omega, k; s) e^{2i\mu BTs}   .  \label{13.2}
\end{eqnarray}

\end{subequations}

We take the differential equation for $ \tilde{\rho}_{12}(t)$ to exemplify the discussion
 of the adiabatic limit of eqs.(\ref{13.0}). At this point we will follow
 closely the references \cite{born,japones}.
 We point out that the adiabatic approximation is not recovered by an $\omega$
 expansion of the terms on the r.h.s. of eqs.(\ref{13.0}).

 The Volterra equation\cite{volterra} obtained from eq.(\ref{13.2}) is

\begin{eqnarray}
\tilde{\rho}_{12} (s)  & = &  \tilde{\rho}_{12} (0)
     + T \int_{0}^{s}  a_{22}(\omega, k; s^\prime) \tilde{\rho}_{12} (s^\prime) ds^\prime
     +T \int_{0}^{s}  a_{21}(\omega, k; s^\prime) e^{2i\mu BTs^\prime}
                                          \tilde{\rho}_{11} (s^\prime) ds^\prime
        +  \nonumber\\
        & + & T\int_{0}^{s}  a_{23}(\omega, k; s^\prime)
                e^{4i\mu BTs^\prime} \tilde{\rho}_{21} (s^\prime)  ds^\prime
            + T \int_{0}^{s}  b_2(\omega, k; s^\prime) e^{2i\mu BTs^\prime}  ds^\prime.
                                                                    \label{13.3}
\end{eqnarray}

In the limit  $T\rightarrow \infty$, the Riemann-Lebesgue
Theorem \cite{churchill} gives that

\begin{equation}
\lim_{\hspace{-.08cm} T \rightarrow \infty} \int_0^s F(s^\prime) e^{i\alpha \mu TB s^\prime}
                   ds^\prime = 0,           \label{14}
\end{equation}

\noindent if  $ F(s^\prime)$ is a piece-wise continuous function
in the  interval [$0,s$] and $\alpha \in  \mathbb{R}$. As a
consequence of this theorem the last three terms on the r.h.s. of
eq.(\ref{13.3}) vanish in the adiabatic limit ($\frac{\omega}{\mu
B}\rightarrow 0 \Rightarrow T \rightarrow \infty $).

Integrating by parts eq.(\ref{14}) we get

\begin{equation}
 \int_0^s F(s^\prime) e^{i\alpha \mu TB s^\prime} ds^\prime =
 \frac{1}{iT} \frac{1}{\alpha \mu B}\left[F(s)   e^{i\alpha \mu TB s} - F(0))   \right]
 - \frac{1}{iT} \frac{1}{\alpha \mu B}
             \int_0^s \frac{d}{ds^\prime}\left( F(s^\prime) \right)
                                                  e^{i\alpha \mu TB s^\prime} ds^\prime.   \label{15}
\end{equation}

From eqs.(\ref{9}) and (\ref{11.2}), the coefficients $a_{2j}$ and
$b_2$ have the general dependence on $\omega$ and $k$:

\begin{subequations} \label{16}

\begin{eqnarray}
a_{21} (\omega, k; s) & = & \omega a_{21}^{(0)} (s) +
  k \left(  \tilde{a}_{21}^{(0)} (s)  + \frac{\omega}{\mu B} A_{21} (s)
    + {\cal O}\left( \left( \frac{\omega}{\mu B} \right)^2 \right) \right), \label {16.1}\\
 a_{22} (\omega, k; s) & = & -i\omega (1- \cos(\theta)) +
  k \left(  \tilde{a}_{22}^{(0)} (s)  + \frac{\omega}{\mu B} A_{22} (s)
     + {\cal O}\left( \left( \frac{\omega}{\mu B} \right)^2 \right)   \right), \label {16.2}\\
  a_{23} (\omega, k; s) & = & k \left(  \tilde{a}_{23}^{(0)} (s)  +
                        \frac{\omega}{\mu B} A_{23} (s)
    + {\cal O}\left( \left( \frac{\omega}{\mu B} \right)^2 \right)   \right) \label {16.3}
\end{eqnarray}

\vspace{-0.5cm}
\noindent and
\vspace{-0.5cm}

\begin{equation}
\hspace{-3cm}
b_2 (\omega, k; s) = \omega b_2^{(0)} (s) + k \left(\tilde{b}_2^{(0)}(s)
        + \frac{\omega}{\mu B} B_2 (s)
       + {\cal O}\left( \left( \frac{\omega}{\mu B} \right)^2 \right)   \right) .
                                                            \label{16.4}
\end{equation}

\end{subequations}

The integrands of all integrals on the r.h.s. of eq.(\ref{13.3}) which
contain the oscillatory function $e^{i \alpha \mu BT s^\prime}$,
with $\alpha =2$ or $4$,  after integration by parts, each of
those integrals has its order in T decreased by one unit, and
acquires a multiplying constant of value $\frac{k}{\mu B}$ or
$\frac{\omega}{\mu B}$. From the conditions satisfied by the two
simultaneous regimes: {\it i}) adiabatic limit
($\frac{\omega}{\mu B} \ll 1$) and {\it ii}) weak coupling limit
($\frac{k}{\mu B} \ll 1$), we can neglect those terms in
comparison to the first two terms on the r.h.s. of
eq.(\ref{13.3}). The two previous inequalities do not impose any
constraint to the ratio $\frac{k}{\omega}$, though. The
differential equation satisfied by $\tilde{\rho}_{12}(s)$ in the
adiabatic limit and weak coupling regime is

\begin{equation}
\frac{d}{ds}\tilde{\rho}_{12} (s) =  T \left[ -i \omega (1- \cos(\theta)) +
   k \tilde{a}_{22}^{(0)}(s) \right]  \tilde{\rho}_{12} (s).      \label{17}
\end{equation}

\noindent The term proportional to $A_{22} (s)$ was dropped,
since it is of higher order in $\left(\frac{\omega}{\mu B}
\right)$. For $k \gg \omega$, the off-diagonal elements of the
density matrix vanish before the external magnetic field
$\vec{B}(t)$ returns to its configuration at $t=0$. The other
uninteresting situation from the point of view of the appearance
of an imaginary correction to the geometric phase is the
condition $ k \ll \omega$ when the dissipation effects can still
be neglected after one period. In this work we discuss the case
$k \sim \omega$ when both terms on the r.h.s. of eq.(\ref{17})
contribute to the time evolution of $\tilde{\rho}_{12} (s)$.

By a similar discussion we obtain  the time equation of
$\tilde{\rho}_{11} (s) $ in the adiabatic and weak coupling
regimes,

\begin{subequations}

\begin{equation}
\frac{d}{ds}\tilde{\rho}_{11} (s) =  T k \tilde{a}_{11}^{(0)}(s) \tilde{\rho}_{11} (s)
   + T \left[\omega b_1^{(0)} (s) + k  \tilde{b}_1^{(0)} (s)  \right].      \label{18.1}
\end{equation}

\noindent From eqs.(\ref{9}) we can affirm that

\begin{eqnarray}
a_{11} (\omega, k; s) &=& k \left( \tilde{a}_{11}^{(0)} (s) +
               \frac{\omega}{\mu B}  A_{11} (s)
   + {\cal O}\left( \left( \frac{\omega}{\mu B} \right)^2 \right)    \right) ,  \label{18.2} \\
b_1 (\omega, k; s) & = & k \left(  \tilde{b}_1^{(0)} (s)+ \frac{\omega}{\mu B} B_1 (s)
   + {\cal O}\left( \left( \frac{\omega}{\mu B} \right)^2 \right)   \right).
                                   \label{18.3}
\end{eqnarray}

\end{subequations}

\vspace{0.5cm}

In order to verify if we can get imaginary phases from eqs.
(\ref{17}) and (\ref{18.1}) due to the coupling of the quantum
system to a dissipative medium, we write the density operator at
$t=0$ in the instantaneous basis of hamiltonian (\ref{4})

\begin{equation}
\rho (0) = \sum_{j, l =1}^2 \; \alpha_{jl}(0) | \phi_j^0 \rangle
\langle \phi_l^0 |. \label{19}
\end{equation}

\noindent We have $ {\bf H}(0) | \phi_j^0 \rangle = E_j(0) | \phi_j^0 \rangle $,
 $ j=1$ and $2$. Being $| \phi_j^0 (t) \rangle$ the time evolution of the eigenvector
$| \phi_j^0 \rangle$, we have

\begin{equation}
\rho (t) = \sum_{j, l =1}^2 \; \alpha_{jl}(t) | \phi_j^0 (t) \rangle
\langle \phi_l^0 (t) |, \label{20}
\end{equation}

\noindent where the time evolution of $| \phi_j^0 (t) \rangle$ is
driven by $ {\bf H}(t)$. Differently from Gamliel and Freed, we include the phases coming from
the unitary evolution and the geometric phase in the dyadic product
$ | \phi_j^0 (t) \rangle \langle \phi_l^0 (t) | $. In the adiabatic
approximation, we get

\begin{equation}
| \phi_j^0 (t) \rangle = e^{ i \;  \gamma_j (t)} \; e^{ -i
\langle E_j(t) \rangle t} | \phi_j ;t  \rangle, \label{21}
\end{equation}

\noindent where $\gamma_j(t)$ is the geometric phase acquired by the
 eigenvector
$|\phi_j^0 \rangle$, $| \phi_j ;t  \rangle$ is the instantaneous
eigenvector of
$ {\bf H} (t)$ ( ${\bf H} (t) | \phi_j ;t  \rangle = E_j(t) | \phi_j ;t
\rangle$ )
and $  \langle E_j(t) \rangle \equiv \frac{1}{t} \int_0^t \; d t^{\prime}
 E_j(t^{\prime})$.

The density operator at any time, in the adiabatic limit is

\begin{equation}
\rho (t) =  \sum_{j,l =1}^2 \; \alpha_{jl}(t) \; e^{ i \; (\gamma_j (t)-
\gamma_l(t)) } \; e^{ - i ( \langle E_j(t) \rangle - \langle E_l(t)
\rangle)  t}  | \phi_j ;t  \rangle  \langle \phi_l ;t | . \label{22}
\end{equation}

From eq.(\ref{22}) we recognize that the phase
 $e^{ i \; (\gamma_j (t)- \gamma_l(t)) } $ in the element $\rho^I_{jl}(t)$ is just the
difference of the geometric phases of the instantaneous eigenstates
$ | \phi_j ;t  \rangle $ and $ | \phi_l ;t  \rangle $ in the absence of
dissipation.

The dynamics of the coefficients $ \alpha_{jl}(t)$ is ruled by the
nonunitary evolution of the quantum system and it is independent
of the particular choice for the instantaneous eigenstates of
the hamiltonian, up to a multiplicative constant\cite{NOTA}. In the
model under consideration (see hamiltonian (\ref{4})), the
eigenvalues $ E_j(t)$, $j=1$ and $2$, are time-independent. It is
simple to get the time equations of $\alpha_{11}(t)$ and
$\alpha_{12}(t)$ from eqs.(\ref{18.1}) and (\ref{17}), respectively

\begin{subequations}\label{23}

\begin{eqnarray}
\frac{d}{dt} \alpha_{11} (t) &= &  k \left( \tilde{a}_{11}^{(0)} (t) \alpha_{11}(t)
     + \tilde{b}_{1}^{(0)} (t)   \right) ,  \label{23.1} \\
\frac{d}{dt} \alpha_{12} (t) &= &  k  \tilde{a}_{22}^{(0)} (t) \alpha_{12}(t).
                                                       \label {23.2}
\end{eqnarray}

\end{subequations}

  The constants $\tilde{a}_{11}^{(0)} (t)$, $\tilde{a}_{22}^{(0)} (t)$
and $\tilde{b}_1^{(0)} (t)$  depend on the particular master
equation that describes the behaviour of the quantum system
interacting with the dissipative medium. In the next  sub-sections
we consider two particular
 interactions of the  two level model with a reservoir: {\it i}) two level model
 in thermal equilibrium  with a reservoir of electromagnetic fields;
 {\it ii}) dephasing process in a two level model.

\subsection{ Adiabatic limit of a two level model in thermal equilibrium }

As discussed before, to incorporate the dissipative effects in the two level
model we introduce the Lindblad superoperator in the diagonal frame.
The master equation of the spin 1/2 model coupled to a reservoir of electromagnetic
 fields in thermal equilibrium in the diagonal
  frame is\cite{lindblad,carmichael}

\begin{eqnarray}
\frac{d}{dt}\rho_D\left( t\right) &=&-{i} \left[
\lambda_1 {\sigma }_{z},\rho_D\left( t\right) \right]
+k\left( \overline{n}+1\right) \left( 2{\sigma }_{-}\rho_D
\left( t\right) {\sigma }_{+}-\rho_D\left( t\right)
{\sigma }_{+}{\sigma }_{-}-{\sigma }_{+}
{\sigma }_{-}\rho_D\left( t\right) \right) + \nonumber \\
%
%
&& + k\overline{n}\left( 2{\sigma }_{+}\rho_D\left( t\right)
{\sigma }_{-}-\rho_D\left( t\right) {\sigma }_{-}
{\sigma }_{+}-{\sigma }_{-}{\sigma }_{+}
\rho_D \left( t\right) \right),
           \label{24}
\end{eqnarray}

\noindent where $k$ is the dissipation constant at zero temperature
and $\bar{n}$ is the average number of excitations of the weakly
coupled thermal oscillators at inverse temperature $\beta$. An important
 requirement for the introduction of  this Lindblad type superoperator
 is that it leads asymptotically to a thermal equilibrium.
In appendix A we give the master equation of this physical process
in the instantaneous basis of hamiltonian (\ref{4}) for arbitrary value of
 $\omega$.

From the master equation in the instantaneous basis of the
hamiltonian we obtain the equations for $\alpha_{11} (t)$ and
 $\alpha_{12} (t)$. These equations in the adiabatic and in the weak coupling limits
become

 \begin{subequations}\label{25}

 \begin{eqnarray}
 \frac{d}{dt}\alpha_{11} (t) & = & -2 k (1+ 2\bar{n})  \alpha_{11} (t)
     + 2 k\bar{n},   \label{25.1}  \\
\frac{d}{dt}\alpha_{12} (t) & = & - k (1+ 2\bar{n})  \alpha_{12} (t).  \label{25.2}
 \end{eqnarray}

 \end{subequations}

\noindent The constant $k$ does not come up on the r.h.s. of
eqs.(\ref{25}) due to the time variation of any classical
parameter that characterizes the reservoir. From its explicit
definition\cite{lindblad,carmichael} it can not be written as:
$f(t)\dot{g}(t)$, where $f(t)$ and $g(t)$ are two regular time
dependent functions. The same
 is true for $\bar{n}$. Therefore the imaginary phase $\chi (t)$
 defined as: $ \alpha_{12} (t) \equiv \alpha_{12} (0) e^{i\chi (t)}$,
 with

 \begin{equation}
  \chi (t) = i \int_{0}^{t} k (1+2\bar{n}) dt^\prime  \label {26}
 \end{equation}

 \noindent is not geometric.

 The solution of eq.(\ref{25.1}) is

 \begin{equation}
 \alpha_{11} (t) = \frac{ \bar{n}}{ 1+2\bar{n}}
 + \left[ \alpha_{11}(0) -  \frac{ \bar{n}}{1+2\bar{n}} \right]
 e^{-2k\int_{0}^{t} (1+2\bar{n}) dt^\prime} .
                                 \label{27}
 \end{equation}

 The exponential decay on the r.h.s. of eq.(\ref{27}) means that the
 population of the instantaneous eigenstates of hamiltonian (\ref{4})
 varies in time and consequently the Adiabatic Theorem is not
 valid for this dissipation mechanism.

Exactly soluble  models are always important checks to
approximation schemes. Eq.(\ref{24}) is exactly solved and the
solutions are

\begin{subequations} \label{sol-ex}

\begin{eqnarray}
\rho_{11}^D(t)&=& \frac{\overline{n}}{2\overline{n}+1} \Big[ 1 -
e^{-2k(2\overline{n}+1)t}\Big] +
\rho_{11}^D(0)e^{-2k(2\overline{n}+1)t}, \label{sol-ex1} \\
\nonumber\\
\nonumber\\
\rho_{12}^D(t)&=& \rho_{12}^D(0)e^{-(2i\lambda_1 +
k(2\overline{n}+1))t}.   \label{sol-ex2}
\end{eqnarray}

\end{subequations}

\noindent It is straightforward to obtain the adiabatic and weak
coupling limit eqs.(\ref{25}) from eqs.(\ref{sol-ex}).

 \subsection{ Dephasing process in two level system}

 Another interesting process well studied in the standard
 textbooks\cite{lindblad,carmichael} is the phase destroying
 process which might appear due to elastic collisions.
 In general, those effects are incorporated in the master equation
 of the two level  model, besides the energy dissipation
 process studied  in subsection 2.1. Since we are
 studying the coupling of the spin 1/2 to a dissipative
 medium in the weak coupling limit, the inclusion of the phase
 destroying process in eq.(\ref{24}) gives corrections to the coefficients
 $a_{ij} (\omega,k;t)$ in eqs.(\ref{25}). Due to the linearity
 of the equations, the new imaginary phases coming from the dephasing
 process are added to the ones obtained previously.
 For the sake of simplicity we study the imaginary phases
 acquired
 by the variables $\alpha_{11}(t)$ and $\alpha_{12}(t)$
 only due to the dephasing process. The master equation
 written in the diagonal frame is

 \begin{equation}
 \frac{d}{dt}\rho_D\left( t\right) =-{i}
 \left[ \lambda_1 {\sigma }_{z},\rho_D\left( t\right) \right]
 + \frac{k}{2}  \left( \sigma_z \rho_D (t) \sigma_z  - \rho_D (t) \right).
                       \label{28}
 \end{equation}

In appendix A we present the master equation of the spin 1/2 with
the dephasing effect included for arbitrary value of the angular
velocity $\omega$ of the external magnetic field. In this
subsection we study the time equations of coefficients
$\alpha_{11} (t)$ and
 $\alpha_{12} (t)$ in the adiabatic and weak coupling regimes. Taking
into account our discussion in section 2, eqs.(\ref{A2}) 
and (\ref{23}) we obtain

\begin{subequations}  \label{29}

\begin{eqnarray}
\frac{d}{dt} \alpha_{11} (t) &  = & 0,  \label{29.1}  \\
\frac{d}{dt} \alpha_{12} (t) &  = & -k \alpha_{12}(t).  \label{29.2}
\end{eqnarray}

\end{subequations}

\noindent  that have the solutions:

\begin{subequations}  \label{30}

\begin{eqnarray}
 \alpha_{11} (t) &  = &\alpha_{11} (0) ,  \label{30.1}  \\
 \alpha_{12} (t) &  = &  \alpha_{12} (0)\; e^{- \int_{0}^{t} k dt^\prime}.   \label{30.2}
\end{eqnarray}

\end{subequations}

\noindent From eq.(\ref{30.1}) we conclude that the Adiabatic
Theorem is valid in this process, since the population at each
quantum state does not vary along the adiabatic process. By
analogous reasons to the ones discussed in subsection 2.1, the
phase in eq.(\ref{30.2}) is not geometric but a time dependent
imaginary phase that destroys the off-diagonal elements of the
density matrix. Eq.(\ref{28}) is exactly solved.
The solutions (\ref{30}) are easily recovered from the exact
solutions when we calculate them in the adiabatic and weak
coupling limits.

\vspace{.5cm}

In order to understand why references
\cite{physlet_a}-\cite{physlet_a2} give an imaginary correction
to the geometric phase and the models studied here do not, we
compare eq.(2.3) of reference \cite{physlet_a} with our
eqs.(\ref{26}) and (\ref{30.2}). The coefficient that multiplies
of variable $C_\lambda (t)$ in eq.(2.3) of reference
\cite{physlet_a} has the form $\langle \theta (t)| \frac{d}{dt} |
\psi (t) \rangle$. Since the vector states depend on a periodic
external parameter $R(t)$, this coefficient, in the adiabatic
approximation, corresponds to a correction to Berry's phase
written as a closed curve in parameter space. The coefficient
$k(1 + 2\overline{n})$ that multiplies $\alpha_{12}(t)$ in
eq.(\ref{25.2}) does not arise from any variation of an external
parameter. The same is true for the coefficient $k$ in
eq.(\ref{29.2}). That is the reason that allows us to claim that
the exponentials acquired from eqs.(\ref{26}) and (\ref{30.2})
have no geometric origin, which means that the suppression terms
are a function of time instead of some path parameter.

\section{Contribution of the geometric phase to  the \hfill \\
                  magnetization}

In order to verify how the geometric phases and the dissipation affect
the physical quantities, we return to the spin $1/2$ model
 in the presence of a reservoir of
electromagnetic fields in thermal equilibrium.
Under the initial condition
 $\rho(0)=|\psi(0)\rangle \langle\psi(0)|$, with $|\psi(0)\rangle =
\cos (\alpha)|+\rangle + \sin(\alpha) |-\rangle$,  in the
adiabatic limit and weak coupling limit we end up with

\begin{subequations} \label{12.todas}

\begin{eqnarray}
\rho_I^{11} (t) &=&
   \frac{ \overline{n}}{ 2\overline{n}+1} +
 \frac{1}{2}\;
\Big( \frac{1}{ 2\overline{n}+1} +
\cos(\theta - 2\alpha) \Big) e^{-2k\left( 2\overline{n}+1\right)t},
       \label{12.0} \\
%
%
&  &  \nonumber\\
%
%
\rho_I^{12} (t) &=&   \frac{1}{2} \, \sin( 2\alpha - \theta)\;e^{-2i\mu Bt}\;
 e^{-k\left( 2\overline{n}+1\right)t}\; e^{-i a \omega(1-\cos(\theta)) t}.
    \label{12}
\end{eqnarray}

\end{subequations}

\noindent We have introduced the Berry's phase tracer $a$ that
helps us identify the contribution of Berry's phase to
physical quantities. At the end, we take the tracer equal to one.

Thus we see that, in this model, dissipation does not affect
Berry's geometrical phase, but only makes it harder to observe
their interference effect: such information is contained in the
off-diagonal terms of the density matrix in the instantaneous
frame, which vanish. By the way, we can observe that the system
considered in this contribution is analogous to the case of the
classical Foucault pendulum where dissipation diminishes the
amplitude, but do not affect the rotation of the oscillation
plane.

 The definition of the geometrical phase can also be given in terms of
 a phase vector as in reference\cite{muller}, where the discussion is
confined to
  pure  states. It is, however, a relatively simple matter to extend the
definition of the
 phase vector for mixed states. In this case we find that 
the loss of coherence  will
 shorten the phase vector in a manner which is completely analogous to what
 happens  to Bloch's vector. This norm reduction, as in the present case,
reflects
  the asymptotic  vanishing
   of off-diagonal density matrix elements\cite{mcn}.

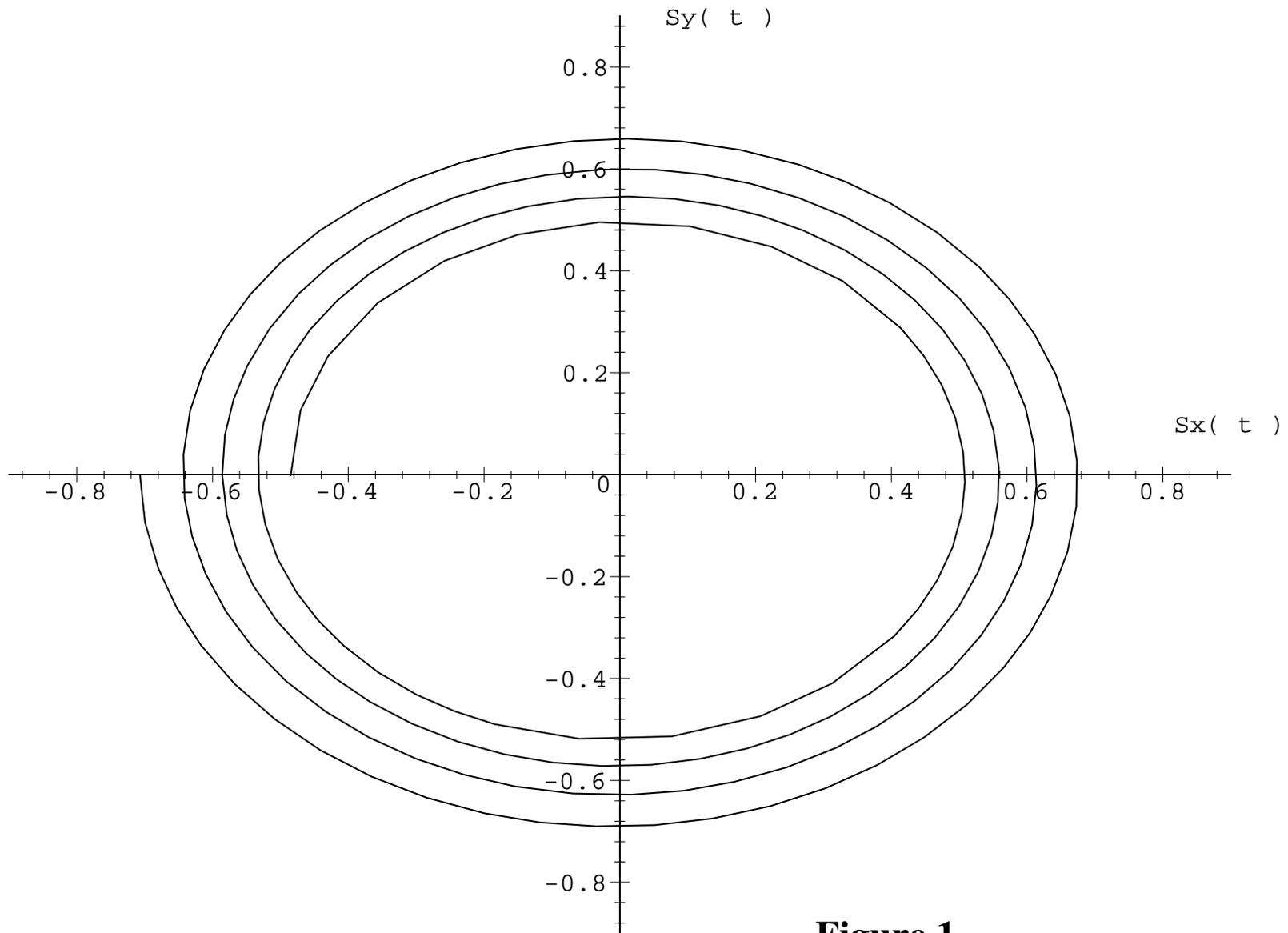
\begin{figure}
\caption{ Plot of the projection of   Bloch's vector
in the $xy$ plane. We take $\alpha =0$,
$\theta= \frac{\pi}{4}$,  $\omega/\mu B = 10^{-3}$ and
$k/\mu B = 10^{-2}$. Time evolution is plotted
in   the interval $t$ = [$ 0, \frac{4 \pi}{\mu B}$].  }
\label{figura01}
\end{figure}

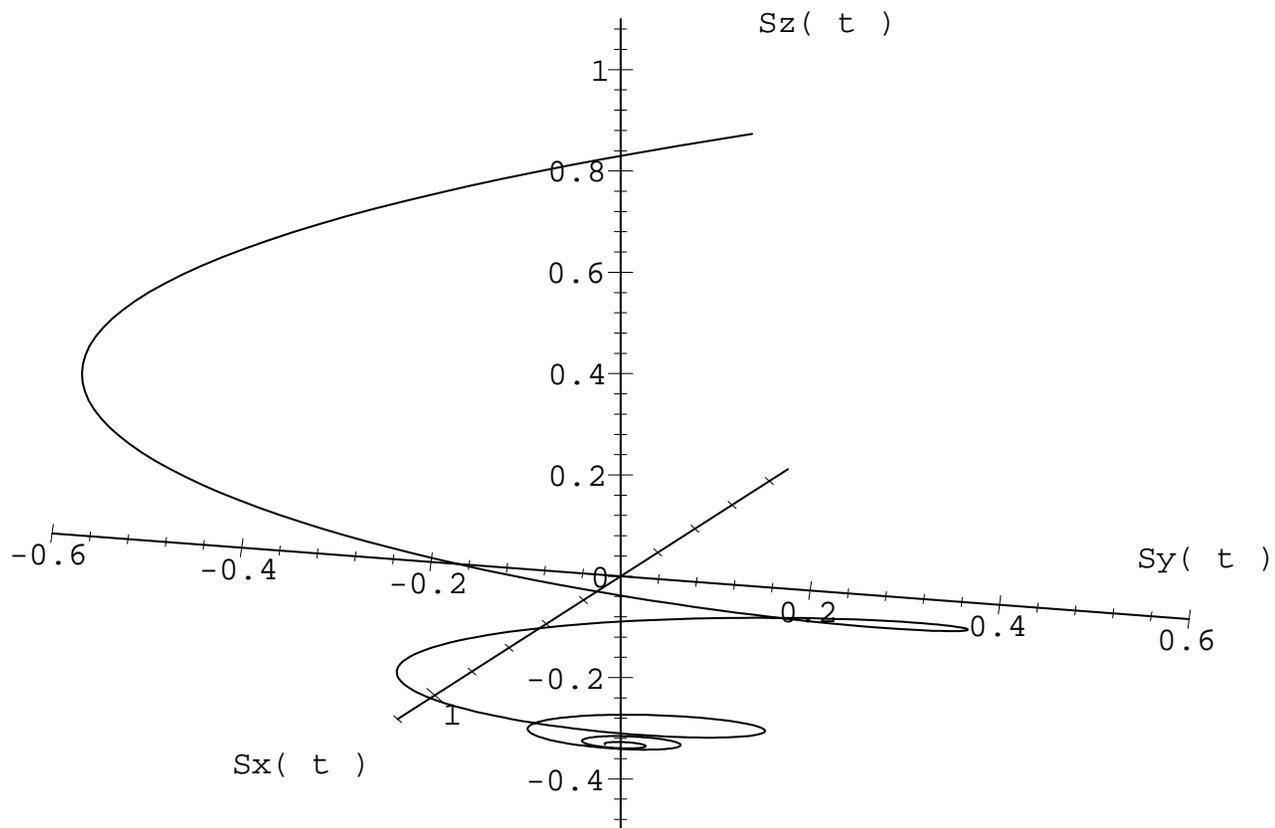
\begin{figure}
\caption{ Curve described by  Bloch's vector in
 space. We take: $\alpha =0$, $\theta= \frac{\pi}{4}$,
$\omega/\mu B = 10^{-3}$ and $k/\mu B = 2. 10^{-4}$. The
vector starts with positive $z$-component  but
ends up on the negative $z$-axis, after a time interval
equal to the period of the external magnetic field.}
\label{figura02}
\end{figure}

Since we are interested on interference effects due to geometric phases,
 Bloch's vector, defined in eq.(\ref{2}), is suitable to provide a graphic
visualization of the density matrix \cite{pra}. From eq.(\ref{3}), we have that
these effects are clearly described by  Bloch's vector in the
instantaneous frame.
In this frame the projection of  Bloch's vector sweeps  the
 $xy$ plane  and makes an angle smaller than
 that of the magnetic field by an amount proportional to the
 solid angle $\Omega(\theta)$, while its length decreases
 exponentially with a time rate of  $k(2\bar{n}+1)$.
We illustrate in figure \ref{figura01} the shrinking of the projection of  Bloch's vector
in the $xy$ plane, due to the presence of dissipation.
Even though we have discussed the case $ k \sim \omega$,
we choose $k/\omega = 10$ in order to show more clearly
the decreasing of this projection in an interval $4\pi/\mu B$.
Shortening in $z$ is faster than in
 the $xy$ plane, causing a time  dependent azimuthal angle.
Figure \ref{figura02} shows the plot
 of the time evolution of  Bloch's vector for $k/\omega= 0.2$.
 For  $t\rightarrow \infty$  Bloch's vector has only
non-zero  $z$-component and
$S_z(t\rightarrow \infty) = - \frac{1}{2\overline{n} +1}$.

For time $t$, such that $t\neq nT$ ($T=2\pi/\omega$ and $n$ is
an integer) the off-diagonal elements of the density matrix $\rho_I(t)$
depends on the chosen condition satisfied by $\langle \phi_i;t| \frac{d}{dt}|
\phi_i;t\rangle$ \cite{nota}. For the sake of
comparison with experiments it is necessary to calculate
 dissipation  and geometrical phases effects on 
 measurable quantities. In this model
the natural candidates are the components of the magnetization
 vector $\langle\vec{m}\rangle (t)$. Let us consider the $z$-component
 of magnetization whose average value has the expression ${\rm Tr}(\rho (t)
 {\bf m_z})$. Since the trace is independent of the particular basis applied
 to calculate it the result is independent of our particular choice of
 $\langle \phi_i; t| \frac{d}{dt} | \phi_i;t \rangle$. In the adiabatic
 approximation and the weak coupling limit, using
 eqs.(\ref{12.todas}), we get the Fourier transform of
 $\langle m_z\rangle (t)$,

 \begin{subequations}
 
\begin{eqnarray}
\langle{\buildrel \sim \over m}_z\rangle (\omega^{\prime}) &=& \hspace{-0.2cm}
\frac{\mu}{\sqrt{2\pi}} \Big[ \frac{\cos(\theta)}{2\overline{n}+1}
\; \pi \delta(\omega^{\prime}) -
i \; \cos(\theta) \Big( \frac{1}{ 2\overline{n}+1} +
\cos(\theta - 2\alpha) \Big) \frac{1}{\omega^{\prime} + 2ki(2\overline{n}
+1)} + \nonumber \\
%
%
 && \hspace{-0.8cm}
- i \; \frac{1}{2}\; \sin(\theta)\sin(\theta - 2\alpha) \;
\Big( \frac{1}{\omega^{\prime} + \Gamma + ki(2\overline{n}
+1)} + \frac{1}{\omega^{\prime} - \Gamma + ki(2\overline{n}
+1)} \Big) \Big],   \nonumber \\
%
%
 &  &     \label{13}
\end{eqnarray}

\noindent where $\alpha$ is given by the initial condition, and the resonant
frequency $\Gamma$ is equal to

\begin{equation}
\Gamma \equiv 2\mu B - a \omega \cos(\theta). \label{gamma}
\end{equation}

\end{subequations}

\noindent The first term of eq.(\ref{13}) corresponds to the constant
component of the magnetic field, and the second one
shows the dissipation effects
 on this field component. The last term on the r.h.s. of the above
 equation displays a real frequency shift which contains the
contribution of the geometrical phase, as can be seen from
eq.(\ref{gamma}), due the presence of the tracer $a$, and a line
broadening caused  only by the dissipative evolution. These
effects on the magnetization agree with the ones derived in reference
\cite{gaitan}  where the path integral formalism was applied.

 The expressions for the other
 components of magnetization are analogous, but somewhat lengthy.

 \section{Conclusions}

In  summary, we have presented an analytical solution  of the adiabatic limit
 of a spin $1/2$ in a precessing magnetic field 
 embedded in a weakly dissipative medium,
 introduced phenomenologically. We consider two distinct 
  nonunitary contributions that were 
 accounted for by a Lindblad type superoperator 
 in the diagonal frame.  We are able to derive  analytical
 expressions for  the geometric and imaginary phases in 
 both cases in the presence of  a weak  dissipation
 in the adiabatic limit without further approximations.

From eqs.(\ref{26}) and (\ref{30.2}) we get that the nature (path
dependent or time dependent) of the imaginary phase acquired by
$\alpha_{12}(t)$ depends on the mechanism that introduces the
dissipation in the quantum system. In both cases that we have studied,
the dissipation is present due to the two level system being in
contact with a  reservoir. The constant $k$ on the r.h.s. of
eq.(\ref{26}) is associated to the time rate of population and
not due to the variation of any external parameter. Consequently
the quantum geometric phase for $k=0$ is not modified by a complex
value.  An analogous argument is valid to explain why the
imaginary phase acquired by the entry $\rho_{12}^I (t)$ in 
the dephasing process is not geometric, either. In this last
model the Adiabatic Theorem continues to be true 
while it is not true anymore for the spin  1/2 coupled
to the electromagnetic fields at thermal equilibrium.
Differently from Ellinas {\it et al.} in
reference\cite{pra} we do not call this complex phase as Berry's
phase. We reserve the name of ``Berry's phase'' only to phases
(real or imaginary) that are path dependent.

Decoherence effects are present and their manifestation is the
shortening of the three components of the Bloch's vector. The
fact that the dissipation effect causes the suppression of
interference patterns due to the geometric phase is not a
particular result for the chosen liouvillian. It is rather
general, stemming from the fact that the dissipation mechanism is
not related to the variation of any set of external periodic
parameters.

\section*{\bf Acknowledgements}

The authors are in debt with M.C. Nemes and J.G. Peixoto de Faria for
useful discussions. The authors thank the referee for bringing to their
attention the interesting model discussed in section 2.2.
A.C.A.P. and M.T.T.  thank E.V. Corr\^ea Silva
for the careful reading of the manuscript.
A.C. Aguiar Pinto thanks CNPq for financial support. M.T. Thomaz  thanks
CNPq for partial financial support.

\section*{Appendix A: Master equations in the instantaneous \hfill \break
           basis of the hamiltonian}

\def\theequation{A.\arabic{equation}}
\setcounter{equation}{0}

The master equation (\ref{24}) in the instantaneous basis of the
hamiltonian for arbitrary value of $\omega$  is

\begin{subequations}\label{A1}
\begin{equation}
\frac{d}{dt}\rho_I\left( t\right) = -{i} \left[
\left( \mu B + \frac{\omega}{2} \right) {\sigma }_{z} -  \frac{\omega}{2} \sigma_n(t) ,
                \rho_I\left( t\right) \right]
                           +k {\cal L}_I \rho_I (t)\label{A1.1}
\end{equation}

\noindent where $\sigma_n(t)$ is given by eq.(\ref{10}) and
\vspace{-0.6cm}

\begin{eqnarray}
{\cal L}_I \rho_I (t) &=& \frac{2\overline{n}+1}{2}\Big\{ -2 \rho_I (t) +
                          (1 - \Lambda^2)\Big[ \sigma_n(t) \rho_I (t) \sigma_n(t) -
                          e^{-2i\omega t} \sigma_+ (t) \rho_I (t) \sigma_+(t) -
                         \nonumber \\
                 \nonumber \\
             &-&  e^{2i\omega t} \sigma_- (t) \rho_I (t) \sigma_-(t) \Big]
             +(1+\Lambda^2)\Big[\sigma_+ (t) \rho_I (t) \sigma_-(t) +
                  \sigma_- (t) \rho_I (t) \sigma_+(t)\Big] -
                            \nonumber \\
             \nonumber \\
             &-& \Lambda\sqrt{1-\Lambda^2}
              \Big[e^{i\omega t} \Big( \sigma_n(t) \rho_I (t) \sigma_-(t)
               + \sigma_- (t) \rho_I (t) \sigma_n(t) \Big) +
                  \nonumber\\
              \nonumber\\
             &+& e^{-i\omega t} \Big( \sigma_+ (t) \rho_I (t) \sigma_n(t) +
                 \sigma_n (t) \rho_I (t) \sigma_+(t) \Big)\Big]\Big\} \; -
                    \nonumber\\
                 \nonumber \\
             &-& \frac{1}{2} \Big\{ \{\rho_I (t), \Lambda \sigma_n(t) +
                    \sqrt{1-\Lambda^2}(e^{-i\omega t}\sigma_+(t) +
                    e^{i\omega t}\sigma_-(t)\} + 2\Lambda\Big[
                    \sigma_+ (t) \rho_I (t) \sigma_-(t) -
                       \nonumber \\
                    \nonumber\\
             &-& \sigma_- (t) \rho_I (t) \sigma_+(t)\Big] + \sqrt{1-\Lambda^2}
             \Big[ e^{-i\omega t} \Big( \sigma_n(t) \rho_I (t) \sigma_+(t) -
             \sigma_+(t) \rho_I (t) \sigma_n(t) \Big) +
                 \nonumber \\
               \nonumber\\
            &+& e^{i\omega t} \Big( \sigma_-(t) \rho_I (t) \sigma_n(t) -
            \sigma_n(t) \rho_I (t) \sigma_-(t) \Big)\Big]\Big\} .
            \label{A1.2}
\end{eqnarray}

\noindent In  eq.(\ref{A1.2}), we define: $\Lambda \equiv \frac{1}{\lambda_1} \Big(
\mu B \cos(\theta) - \frac{\omega}{2}\Big)$  and

\begin{equation}
\sigma_+(t) = e^{i\omega t} \Big[ \frac{\sin(\theta)}{2} \sigma_z +
                   (\cos(\frac{\theta}{2}))^2 e^{-i\omega t}\sigma_+ -
                   (\sin(\frac{\theta}{2}))^2 e^{i\omega t} \sigma_- \Big]
            \label{A1.3}
\end{equation}

\end{subequations}

\noindent  and   $\sigma_-(t) \equiv \Big(\sigma_+(t)\Big)^\dagger$.

\vspace{2cm}

The master equation (\ref{28}) in the instantaneous basis of the
hamiltonian for arbitrary value of $\omega$ is

\begin{subequations}\label{A2}
\begin{equation}
\frac{d}{dt}\rho_I\left( t\right) = -{i} \left[
\left( \mu B + \frac{\omega}{2} \right) {\sigma }_{z} -  \frac{\omega}{2} \sigma_n(t) ,
                \rho_I\left( t\right) \right]
                           +\frac{k}{2} {\cal L}_I \rho_I (t)\label{A2.1}
\end{equation}

\noindent where
\vspace{-0.6cm}

\begin{eqnarray}
{\cal L}_I \rho_I (t) &=& \Lambda^2 \sigma_n(t) \rho_I (t) \sigma_n(t) +
                          \Lambda\sqrt{1-\Lambda^2} \Big[
                          e^{-i\omega t} \Big( \sigma_n(t) \rho_I (t) \sigma_+(t) +
                          \sigma_+(t) \rho_I (t) \sigma_n(t) \Big) +
                              \nonumber \\
                                 \nonumber \\
                    &+& e^{i\omega t} \Big( \sigma_n(t) \rho_I (t) \sigma_-(t) +
                        \sigma_-(t) \rho_I (t) \sigma_n(t) \Big) \Big] +
                        (1-\Lambda^2)\Big[
                        e^{-2i\omega t} \sigma_+(t) \rho_I (t) \sigma_+(t) +
                           \nonumber \\
                              \nonumber\\
                    &+& e^{2i\omega t} \sigma_-(t) \rho_I (t) \sigma_-(t) +
                        \sigma_+(t) \rho_I (t) \sigma_-(t) +
                        \sigma_-(t) \rho_I (t) \sigma_+(t) \Big] - \rho_I(t).
                    \label{A2.2}
\end{eqnarray}

\end{subequations}

\newpage



\begin{thebibliography}{99}
\bibitem{nota} Like the case of no dissipation, the off-diagonal
elements of the density matrix are not uniquely defined unless we
impose a condition on $\langle \phi_i;t|\Big(\frac{d}{dt}|
\phi_i;t\rangle\Big)$, $i=1$ and $2$. In order to get eq.(\ref{8}), we
impose the conditions:  $\langle \phi_1;t|\Big(\frac{d}{dt}|
\phi_1;t\rangle \Big) = i\omega\sin^2(\frac{\theta}{2})$ and
$\langle \phi_2;t|\Big(\frac{d}{dt}| \phi_2 ;t\rangle\Big) = -
i\omega\sin^2(\frac{\theta}{2})$.
\bibitem{NOTA} If we make another choice for the instantaneous
eigenstates of hamiltonian, $|\Phi_i;t\rangle = e^{i
\beta_i(t)}|\phi_i;t\rangle$, $i=1$ and $2$, the dynamics of the
coefficient $\alpha_{jl}^{(\beta)} (t)$ in the new basis is related
to to the dynamics of $\alpha_{jl} (t)$ (from eqs.
(\ref{23})) by $\frac{d \alpha_{jl}^{(\beta)}(t)}{dt} =
e^{-i(\beta_j(0)-\beta_l(0))} \frac{d \alpha_{jl}(t)}{dt}$.
\end{thebibliography}
\end{document}